# FROM DATA TO DESIGN: RANDOM FOREST REGRESSION MODEL FOR PREDICTING MECHANICAL PROPERTIES OF ALLOY STEEL

**Samjukta Sinha1\*, Prabhat Das2,3**
[1]Department of Metallurgy and Materials Engineering, Indian Institute of Engineering Science andTechnology, Shibpur, Howrah-711103,WestBengal, India
[2]Department of Information Technology, School of Computing Sciences, The Assam Kaziranga University, Jorhat-785006, Assam, India
3Department of Computer Science and Engineering, School of Engineering and Technology, Adamas University, Kolkata-700126, West Bengal, India
\*samjukta.sinha@gmail.com

**Abstract :** This study investigates the application of Random Forest Regression for predicting mechanical properties of alloy steel-Elongation, Tensile Strength, and Yield Strength-from material composition features including Iron (Fe), Chromium (Cr), Nickel (Ni), Manganese (Mn), Silicon (Si), Copper (Cu), Carbon (C), and deformation percentage during cold rolling. Utilizing a dataset comprising these features, we trained and evaluated the Random Forest model, achieving high predictive performance as evidenced by $R^2$ scores and Mean Squared Errors (MSE). The results demonstrate the model's efficacy in providing accurate predictions, which is validated through various performance metrics including residual plots and learning curves. The findings underscore the potential of ensemble learning techniques in enhancing material property predictions, with implications for industrial applications in material science.

**Introduction :**

Alloy steel is a versatile and widely used category of steel that is defined by the deliberate addition of alloying elements such as manganese, chromium, nickel, molybdenum, vanadium, and silicon, among others. These elements are added to the base iron-carbon mixture to enhance specific mechanical properties like strength, toughness, hardness, ductility, and corrosion resistance, making alloy steels suitable for a broad range of industrial applications [1]. The basic definition of steel is an alloy of iron and carbon, where carbon typically constitutes upto 2.1% of the composition [2]. However, when additional elements are introduced, the resulting alloy steels exhibit properties that are far superior to those of plain carbon steel, enabling them to meet the demands of various high-performance applications .

In materials science, the "banana diagram" as shown in Figure 1 is a useful graphical tool that illustrates the trade-offs between different mechanical properties, most notably strength and ductility. The curve on this diagram typically shows that as strength increases, ductility decreases, forming a banana-shaped curve. This trade-off is a critical consideration when designing alloy steels, as engineers must find the optimal balance between these properties to meet the specific requirements of different applications [4]. For example, in the automotive industry, parts such as engine components need to be strong enough to withstand high stresses, while other parts, like the vehicle body, must be ductile enough to absorb impact energy in the event of a crash [5].

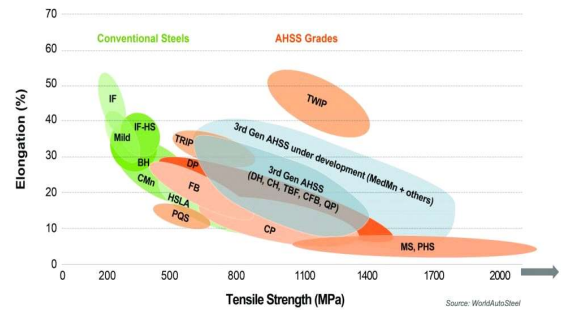

**Figure 1:** Steel strength ductility diagram [6]

Key mechanical properties like tensile strength, yield strength, and elongation are critical for evaluating steel performance. Steel strength improves through plastic deformation processes like rolling or forging, which introduce dislocations in the crystal structure [7]. As deformation continues, dislocation density rises, causing strain hardening, making further dislocation movement more difficult and increasing material strength [8]. The Hall-Petch relationship further explains that yield strength increases as grain size decreases, with strength being inversely proportional to the square root of the grain size.

$$\sigma_y = \sigma_0 + k \cdot d^{\frac{-1}{2}}$$

where:
- $\sigma_y$ is the yield strength,
- $\sigma_0$ is a constant (friction stress),
- 'k' is the Hall-Petch slope (a material constant related to the hardening mechanism),
- 'd' is the average grain diameter [9].

As the grain size decreases (which can be achieved through deformation processes such as cold working), the yield strength increases because smaller grains provide more grain boundaries, which act as barriers to dislocation movement [10].

Alloying is another method used to enhance the strength of steel, primarily through **solid solution hardening.** This occurs when alloying elements are added to the steel, creating a solid solution where the atoms of the alloying element replace or occupy interstitial positions within the iron lattice. The presence of these atoms distorts the crystal lattice, creating stress fields that impede dislocation movement, thereby increasing the strength of the steel [11]. For example, adding chromium to steel introduces atoms of different sizes into the iron lattice, creating a more complex crystal structure that is harder for dislocations to move through [12].

Alloy steels are essential in the automotive sector for their customizable properties. Chromium and copper enhance corrosion resistance, while molybdenum improves high-temperature strength, ideal for engine parts [13-15]. Nickel boosts toughness for stress and low temperatures [16], and manganese plus silicon increase hardenability and tensile strength [17]. Higher carbon content improves hardness and strength but reduces ductility [18]. This customization helps in producing lighter, stronger, and more durable vehicle parts.

Traditionally, alloy steel tailoring relied on empirical methods and extensive testing. Machine learning (ML), a branch of AI, has transformed this by analyzing large datasets to predict mechanical properties based on alloy composition and heat treatment [19]. ML accelerates development and enables discovery of new alloys with desirable properties without extensive physical testing.

In ML, supervised learning is one of the most common techniques, where the model is trained on a labeled dataset [20]. An advanced form of ML, known as ensemble learning, combines the predictions from multiple models to improve accuracy. The idea behind ensemble methods is that by combining the strengths of different models, the ensemble can produce better predictions than any single model. When applying ensemble learning to predict the mechanical properties of alloy steel, multiple individual models (e.g., decision trees, linear regression models) are trained on the same dataset. This approach reduces the risk of overfitting and improves the model's generalization ability [21].

Machine learning models are evaluated using statistical metrics such as the R-squared ($R^2$) score and the Mean Square Error (MSE). The $R^2$ score is a statistical measure that indicates how well the predicted values from a model match the actual data. It is defined by the formula:

$$R^2 = 1 - \frac{\sum_{i=1}^{n}(y_i - \hat{y}_i)^2}{\sum_{i=1}^{n}(y_i - \bar{y}_i)^2}$$

Where, (a) $\sum_{i=1}^{n}(y_i - \hat{y}_i)^2$ is the sum of squared residuals.

(b) $\sum_{i=1}^{n}(y_i - \bar{y}_i)^2$ is the total sum of squares, where $\bar{y}$ is the mean of the actual values.

(c) The $R^2$ Score value ranges from 0 to 1, where 1 indicates a perfect fit of the model to the data.

An $R^2$ score close to 1 indicates that the model explains most of the variance in the data, meaning the predictions are highly accurate.

The Mean Square Error (MSE) is another key metric used to evaluate the accuracy of an ML model. It is the average of the squares of the differences between predicted and actual values and is given by the formula:

$$MSE = \frac{1}{n}\sum_{i=1}^{n}(y_i - \hat{y}_i)^2$$

Where, (a) $n$ is the number of data points

(b) $y_i$ is the actual value

(c) $\hat{y}_i$ is the predicted value.

A lower MSE indicates that the predictions are close to the actual values, which is crucial for ensuring the model's accuracy [22].

The model with the highest $R^2$ score and lowest MSE is the most accurate predictor, guiding the design of steel alloys with optimal properties. Machine learning revolutionizes materials science by enabling precise prediction of alloy properties, leading to stronger, lighter, and more durable materials. This benefits industries such as automotive, aerospace, construction, and energy, where advanced materials are crucial. As ML techniques advance, their role in metallurgy will become essential, driving innovation in steel-based products.

This study models the impact of alloy steel composition and cold rolling deformation on mechanical properties using ensemble machine learning techniques. By combining multiple models, the approach aims for superior accuracy. The model with the best performance, indicated by high $R^2$ and low MSE, is used for predicting tensile strength, yield strength, and elongation, ensuring reliable and optimized predictions for the alloy steel dataset.

## 2. Dataset and Modelling :

The dataset, compiled from various literature, includes around 300 experimental data points with alloy steel composition (74-83 wt% Fe, 0-9 wt% Cr, 0.5-10 wt% Mn, 0-15 wt% Ni, 0-6 wt% Si, 0-2 wt% Cu, 0-3.7 wt% C) and deformation percentage as input parameters. The outputs are tensile strength, yield strength, and % elongation. An ensemble classifier was used, combining the votes of four ML models: (1) Linear Regression, (2) Random Forest Regressor, (3) Support Vector Regressor, and (4) Gradient Boosting Regressor for accurate predictions (Fig. 2).

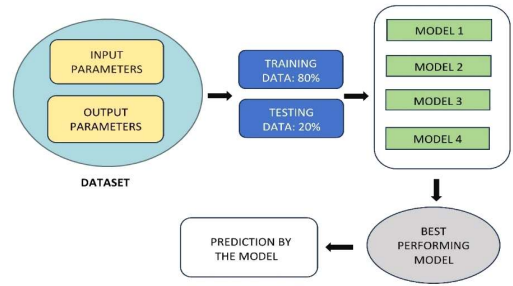

**Figure 2:** Pictorial Representation of voting based ensemble machine learning technique.

## Experiment :
### 3.1 Dataset:

The dataset used for this study contains features related to composition of Iron (Fe), Chromium (Cr), Nickel (Ni), Manganese (Mn), Silicon (Si), Copper (Cu), Carbon (C), and deformation percentage during cold rolling. The target variables are Elongation, Tensile Strength, and Yield Strength.

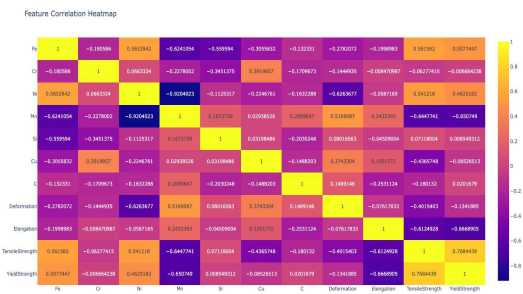

**Figure 3:** Feature correlation Heatmap

Each cell in the heatmap (Figure 3) shows the correlation coefficient between two variables, with color intensity representing the strength and direction of the correlation. A coefficient of +1 denotes a perfect positive correlation (both variables increase together), while -1 indicates a perfect negative correlation (one variable increases as the other decreases) [23].

The heatmap highlights significant relationships, such as strong correlations between Ni and tensile/yield strength and Mn with elongation, emphasizing their roles in material strength and ductility. Negative correlations, like Cr and Si with elongation and Mn with tensile/yield strength, suggest trade-offs were increasing one element may decrease another. Analyzing these relationships helps optimize material composition for desired properties.

The dataset, extracted from a CSV file, includes element contents (Fe, Cr, Ni, Mn, Si, Cu, C) and percentage deformation. Target variables are elongation, tensile strength, and yield strength. The data was split into training (80%) and testing (20%) sets using Scikit-learn's train_test_split function to train and evaluate the model. Various machine learning regressor algorithms were applied, with performance summarized in Table 1.

**Table 1:** Evaluation metrics ($R^2$, MSE and Accuracy) of each Model

| Model | $R^2$ Score | MSE | Accuracy |
|---|---|---|---|
| Linear Regression | 0.731782 | 4661.505696 | 0.731782 |
| Random Forest Regressor | 0.928557 | 2278.976453 | 0.928557 |
| Support Vector Regressor | -0.002361 | 25008.742173 | -0.002361 |
| Gradient Boosting Regressor | 0.924486 | 2403.618269 | 0.924486 |

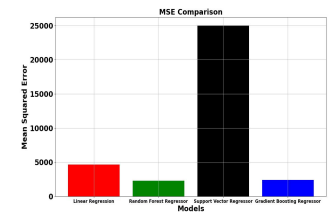 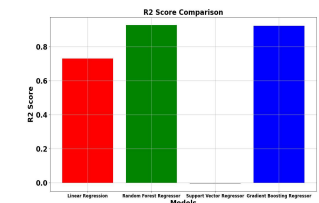

**Figure 4:** Graphical representation of mean square error of different models

**Figure 5:** Graphical representation of $R^2$ score comparison of different models

Table 1 presents performance metrics for various regression models in predicting mechanical properties. The Random Forest Regressor excels with the highest R² score (0.9286) and the lowest Mean Squared Error (2278.98), demonstrating superior accuracy and minimal error. Its strength lies in its robustness against overfitting and its capability to manage complex, non-linear relationships. Therefore, the Random Forest Regressor constructs multiple decision trees and averages their predictions, and hence was chosen for prediction due to its effectiveness in handling complex datasets and improving prediction accuracy. We initialized the model with 200 decision trees (`n_estimators=200`) and trained on the training dataset. After training, the model was used to predict the target variables on the test dataset.

## 4. Results and Discussion :

The performance of the Random forest regressor model was evaluated using the R² (coefficient of determination) and Mean Squared Error (MSE) metrics for each target variable which is represented in Table 2.

**Table 2:** Evaluation metrics (R² and MSE) for each target variable

| Target Variable | R² Score | MSE |
|---|---|---|
| Elongation | 0.9437 | 1.6568 |
| Tensile Strength | 0.9852 | 433.3525 |
| Yield Strength | 0.8544 | 6587.3567 |
| Average R² | 0.9278 | - |

The high R² scores indicate that the model explains a significant portion of the variance in the target variables, particularly for tensile strength, which has an R² of 0.9852, suggesting a very strong predictive capability. The MSE values, although varying in magnitude, are reflective of the scale of the respective target variables.

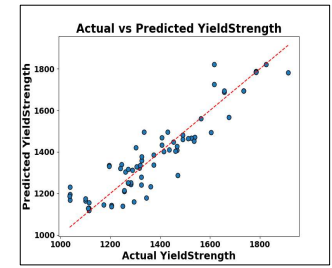 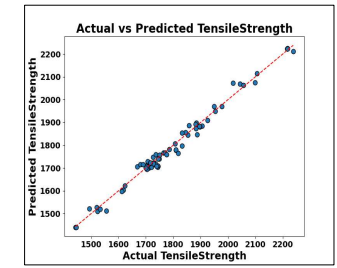

**Figure 6:** Actual yield strength of a material against the predicted yield strength.

**Figure 7:** Actual tensile strength of a material against the predicted tensile

Figures 6 to 9 show a strong correlation between actual and predicted values, with data points clustering around the diagonal line, confirming the model's effectiveness. Some scatter and outliers suggest minor inaccuracies or unaccounted factors. The training score is consistently high (close to 1), while the validation score starts lower but improves to about 0.85 with larger training sets. This indicates initial overfitting with smaller datasets, but better generalization with larger sets.

The Random Forest Regressor effectively predicts Elongation, Tensile Strength, and Yield Strength from alloy steel composition and deformation percentage. High R² scores and low MSE values show the model's accuracy and reliability, outperforming traditional methods.

Overall, the Random Forest Regressor is a reliable tool for predicting material properties and can be used effectively for similar tasks.

## Conclusion :

The implications of this study extend to practical applications in the field of material science and engineering, where adequately precise predictions of mechanical properties can lead to optimized material selection and improved design processes. Furthermore, these advanced predictive capabilities facilitate the design of new materials with tailored properties, leading to innovations in product development and manufacturing processes. The use of machine learning not only streamlines the material design process but also enhances the precision of simulations and optimizations, ultimately contributing to more efficient and effective engineering solutions. This integration of cutting-edge technology into traditional material science practices underscores the transformative potential of machine learning in advancing the field and achieving higher standards of performance and reliability in engineered materials.

<div style="text-align:center">** ** **</div>